# Viscoelastic-Viscoplastic Strength Model for PBX

Roman Kositski
Rafael Ltd. POB 2250, Haifa, 3102102 Israel
kositski@gmail.com

**Abstract.** We present a material (strength) model for describing the deformation of PBX under quasistatic mechanical and thermal loadings. It is a viscoelastic-viscoplastic model incorporating the effects of pressure and temperature on the strength of PBX. The model is calibrated using uniaxial tension and compression tests. Using FEM simulations in Ls-Dyna, we validated the model with creep and relaxation tests. To further validate the model, we performed an experiment with thermally loaded, biaxially strained, sample and measured its deformation using Digital Image Correlation method.

## Introduction

Polymer (or Plastic) bonded explosive (PBX) are common in a variety of defense applications. PBX is used in Reactive and Active Armor protection systems, Shaped Charge antitank warheads, as well as various other systems. The main objective of the explosive in a system is to detonate when needed and rapidly release chemical energy. In many applications, the explosive also has to sustain mechanical and thermal loads during its life cycle. As new weapon development relies more and more on numerical modeling, reliable material modeling is needed to calculate the strength and deformation of PBX as much as it is needed for any other structural material.

For normal operation conditions, most structural materials, e.g. plastics and metals, can be modeled using relatively simple elastic or elasto-plastic strength models. However, under the same conditions the mechanical behavior of most PBXs exhibit temperature and rate dependence [1], viscous effects such as creep and relaxation [2], and different strength values under compression and tension [3]. This, rather unique, set of properties requires a special material model [4] to be used in finite element method (FEM) codes.

Several models were recently proposed to model PBX at quasistatic loading [4–7]. While those models capture some or all the required features of PBX deformation as described above, they are complex to implement in a FEM code and difficult to calibrate.

In this work, we propose a material model which is relatively simple to implement and calibrate. We also present a validation experiment in which the PBX sample is thermally loaded to achieve a state of bi-axial tension.

## Material Model

The presented material model is an extension of the Phases Model (PHM) which was proposed by Partom and Schanin [8] for a one dimensional (1D) case and expanded to a 2D finite difference code by Keren et al. [9]. The material model is based on the decomposition of the stress tensor into a hydrostatic part and a deviatoric part.

$$\sigma_{ij} = S_{ij} - \frac{1}{3}\sigma_{ii}\delta_{ij} = S_{ij} - P\delta_{ij}, \qquad (1)$$

where $S_{ij}$ denotes the deviatoric stress tensor, $P$ denotes the pressure, defined positive in compression, and $\delta_{ij}$ is the Kronecker delta. The pressure is calculated using a simple linear Equation of State (EOS) with the bulk modulus $K$.

$$P = -K\varepsilon_v \qquad (2)$$

where $\varepsilon_v$ is the volumetric strain (positive in tension).

The deviatoric stress tensor is calculated on the basis of a generalized Maxwell model with a set of linear springs and non-linear dashpots, connected in parallel as shown schematically in Figure 1a. Both the springs and dashpots are temperature dependent. Following the nomenclature of [8], each Maxwell element is named a "Phase". The deviatoric stress tensor is a sum of stresses from all phases

$$S_{ij} = \sum A^k S_{ij}^k \qquad (3)$$

where $A^k$ is the weight of the $k^{th}$ phase, such that

$$\sum A^k = 1. \qquad (4)$$

The stress rate in every phase can be expressed as a set of rate equations:

$$\dot{S}_{ij}^k = 2G(\dot{d}_{ij}^{elastic}) = 2G(\dot{d}_{ij} - \dot{d}_{ij}^{vis}) \qquad (5)$$

where $\dot{d}_{ij}$ is the total deviatoric strain rate tensor and $\dot{d}_{ij}^{visco}$ is the viscous strain rate tensor (which by definition has only a deviatoric component) and $G$ is the shear modulus. Based on the Prandtl-Reuss approach, we assume that the principal plastic strain increments are proportional to the principal deviatoric stresses, so that the viscous strain rate tensor is

$$\dot{d}_{ij}^{visco} = \frac{3}{2}\dot{\varepsilon}^v \frac{S_{ij}}{\sigma_{eff}} \qquad (6)$$

where the effective stress $\sigma_{eff}$ is defined as

$$\sigma_{eff} = \sqrt{\frac{3}{2}S_{ij}S_{ij}}. \qquad (7)$$

Finally, we define "flow curves" of the/ dashpots as

$$\dot{\varepsilon}^v = f(\sigma_{eff}, T, P) \qquad (8)$$

A flow curve can be any monotonically increasing function and we implement them as a look-up table. A set of flow curves is shown in Figure 1b. In total, to solve for the deviatoric stress we need to integrate $N$ sets of the above equations where $N$ is the number of phases.

In their paper [8], Partom and Schanin presented a method for calibrating the described "Phases Model" based on uniaxial tension tests performed at several constant strain rates. For the sake of brevity, we do not repeat this description here. To incorporate the effect of temperature, we follow the calibration procedure at several temperatures resulting in temperature dependence of all parameters (K, G, and flow curves).

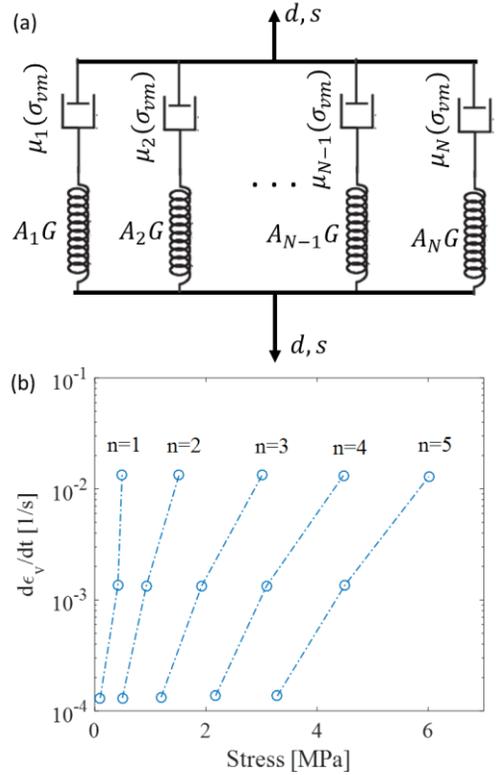

Figure 1 - (a) A schematic representation of the Maxwell elements modeling the deviatoric response. (b) An example of a set of 5 flow curves for a given temperature.

**Pressure effects**

The strength of PBX is very sensitive to the stress state [10] and it is very asymmetric in tension vs. compression. One way to incorporate this effect is to use pressure as a governing parameter. An example of the extreme effect of the loading direction can be clearly seen in Figure 2.

Based on the stress strain curves in Figure 2 it seems that the pressure does not affect the initial elastic modulus (at least at low pressures), so we assign the pressure dependence in the PHM only to the dashpot (or viscous) elements. The pressure "shifts" the flow curves, for a given temperature

$$\dot{\varepsilon}^v = f(\sigma_{eff} - \alpha P) \qquad (9)$$

where $\alpha$ is a fitting parameter. As $f$ is a monotonically increasing function, increasing the pressure (e.g. compression) results in a lower viscos strain and as a result a higher elastic strain and a higher stress.

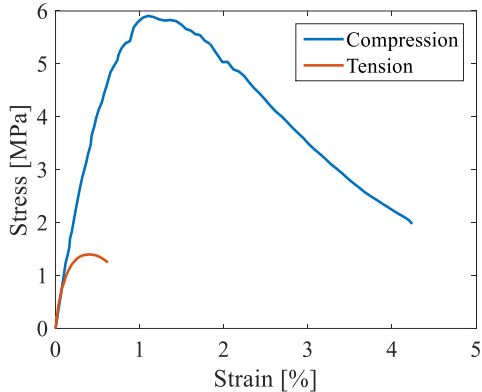

Figure 2 - Stress strain curves obtained from uniaxial tension and compression tests on PBX from the same material batch.

**Calibration experiments**

The material chosen for this study is a pressed PBX composed of HMX and a fluoropolymer binder. To calibrate the PHM we performed uniaxial tension and compression tests at temperatures ranging from 0°C to 50°C. The tests were performed on a uniaxial tension/compression machine with cross-head speeds of 0.5, 0.05 and 0.01 mm/min, resulting in strain rates of about $10^{-4}$, $10^{-5}$ and $2 \times 10^{-6}$ s$^{-1}$ respectively. The measured stress strain curves for tension and compression at several temperatures are shown in Figure 3.

The stress strain curves, both for tension and compression of the chosen PBX, exhibit a softening response rather than brittle fracture. A similar response was observed and studied by other researchers [3], and it is believed to be the result of damage evolution inside the material. The proposed model does not account for this effect and all our stress strain curves were truncated once maximum stress is reached, and this is also the validity range of the current model.

Using the uniaxial tension stress-strain curves, the flow curves and phase weights ($A^k$) were calibrated as described in [8]. The $\alpha$ parameter for the pressure dependence was then calibrated and validated on the basis of the compression experiments.

**Numerical Modeling**

The Phase Model was implemented as a material model in Ls-Dyna for solid elements both in 2D and 3D solvers. The model is suited for an explicit integration scheme. As such, very small time-steps during problem solution may occur. To overcome this limitation of the explicit scheme in modeling quasi-static loading scenarios, we use the artificial mass scaling technique implemented in Ls-Dyna [11]. Using this method, we can perform calculations with time steps of 0.1 – 1 sec that result in reasonable solution time for a typical loading scenario of several minutes or even hours.

The simulated stress and strain curves for tension and compression are superimposed on the experimental curves in Figure 3. The results of the tension simulations are in good agreement with the experiments. For compression, the model over-estimates the stress at higher strains. This may be improved by adding additional phases (all results reported here use 5 phases), or by further modifying the pressure dependence.

**Validation experiments and modeling**

Although the calibration of the model uses discrete values from the stress strain curves, these curves are still seen as calibration curves and we wish to perform other experiments and calculations to validate the model.

Tensile creep tests

Tensile creep tests were performed using a "dead-load" method. The experiments were performed at 25°C and -5°C with two loads at each temperature. The measured and simulated strain histories are shown in Figure 4. Since damage is not modeled, the model cannot capture the tertiary creep seen in the 1.2 MPa -5°C experiment after about 80 minutes.

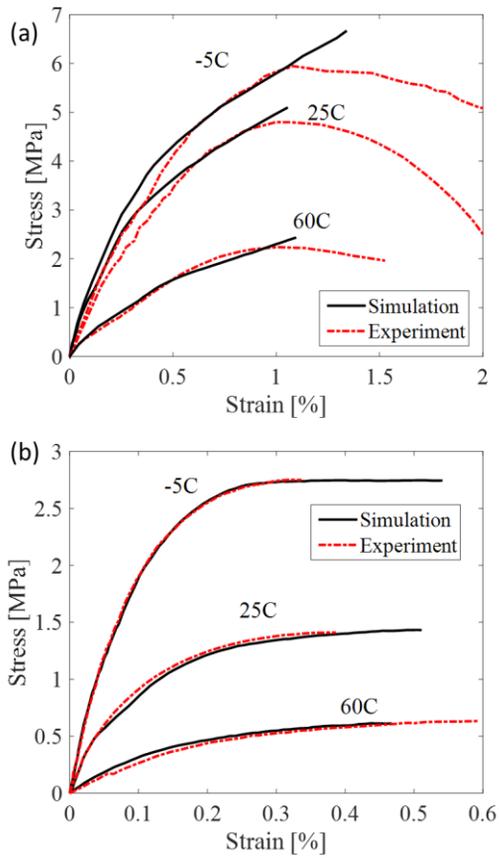

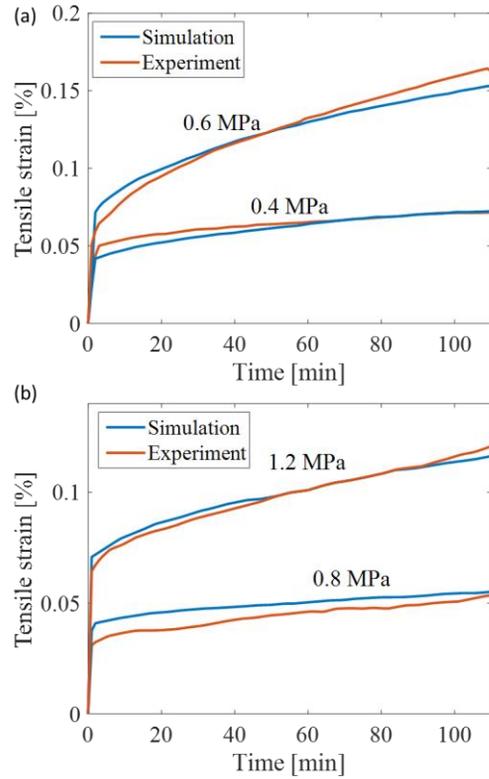

Figure 4 - Tension creep tests and simulations at (a) 25°C and (b) -5°C.

Figure 3 -Stress strain curves in (a) compression and (b) tension at a loading rate of 0.5 mm/min, obtained at -5°C, 25°C and 60°C. Experiments and simulations.

Stress relaxation tests

We performed stress relaxation tests by loading a specimen in tension to a predefined stress (0.4 and 0.6 MPa) at a rate of 0.5 mm/min and holding the strain constant (measured with extensometer). During the experiment, the stress is monitored over time. Relaxation curves with different initial stress are plotted in Figure 5. In both cases, the material model reproduces the measured curves reasonably well.

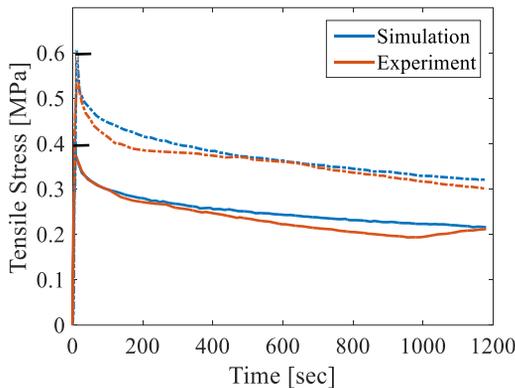

Figure 5 - Stress relaxation from a tensile test to initial stress of 0.4 and 0.6 MPa.

Bi-Axial thermal loading.

Up until now we only discussed uniaxial experiments and simulations at constant temperatures. As the material model is implemented in a full 3D framework we wish to validate it in a non-uniaxial-stress setting. We confined a round PBX sample in a steel ring by gluing it. The thermal expansion coefficient of the PBX is about 5 times higher than that of steel [12]. As the sample and ring are cooled, the PBX tends to contract while the steel ring restrains it thru the glued bond, effectively straining it radially, and resulting in a biaxial stress state. The sample and steel ring are shown in Figure 6. The test is performed by cooling the sample at a constant rate and measuring the strains on the sample (at the gauge section) using a video extensometer. A photograph of the sample with a speckle pattern on it and the location of the "placed" virtual strain gauges are shown in Figure 7. The measured temperature during the test was imposed as boundary conditions in the numerical simulation. We used the built-in thermal solver of Ls-Dyna [13] so both the thermal and mechanical problems were solved simultaneously. The temperature profile, as well as the measured and calculated mechanical strains, are plotted in Figure 8. During the first loading cycle, the simulation closely follows the experimental results. During the unloading stage (temperature increase) and second cooling cycle, the model starts to deviate from the experiment. This residual strain is the direct result of a plastic strain accumulating in the simulation as the stresses approach the maximum stress from the strain stress curves. However, in the experiment, the sample seems to fully recover all strains. This can be explained by the evolution of damage during the loading that effectively softens the material but keeps the strains elastic [2].

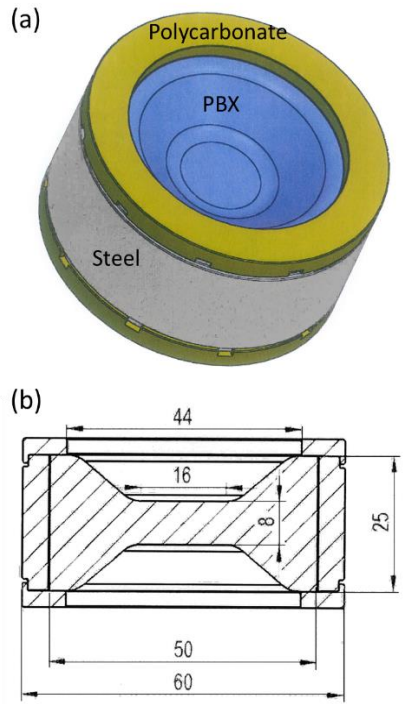

Figure 6 - (a) CAD model of the biaxial experiment. (b) A drawing with dimensions (mm).

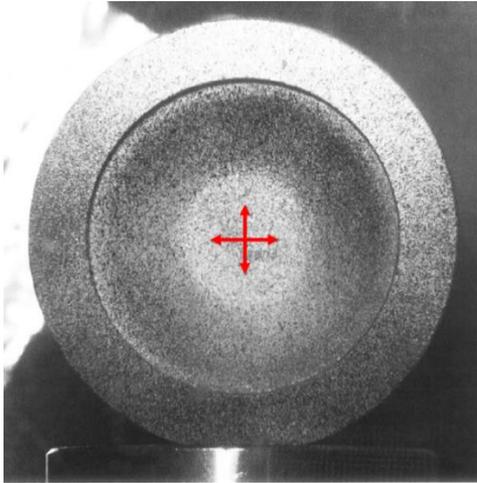

Figure 7 - The sample covered with speckles for the DIC analysis with virtual strain gauges placed on the gauge section.

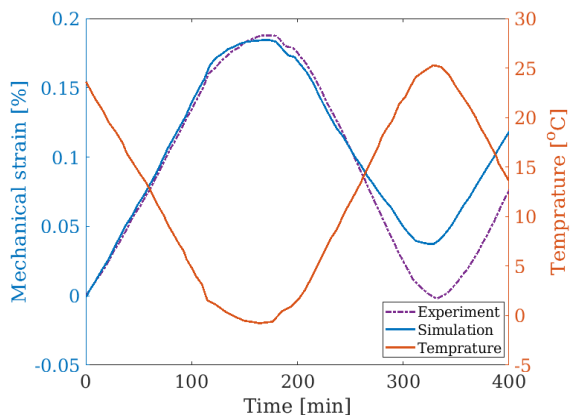

Figure 8 - Measured and calculated mechanical strain and the external boundary temperature.

## Conclusions and outlook

The presented model uses the approach of Partom and Schanin [8] to describe the viscoelastic-viscoplastic temperature and pressure dependent strength of a pressed PBX. This approach was successfully implemented in a modern FEM code and calibrated with tension and compression experiments. While the presented model is successfully used to describe various complex systems and loading scenarios, it lacks a description of damage evolution in the material. Damage, such as deboning between the explosive crystals and the polymer matrix is believed to be a major factor affecting the deformation of PBXs [14]. We are currently working on incorporating a damage parameter into the PHM which will result in "softening" the springs and effectively weakening the material while maintaining fully reversible strains. We believe that this modification to the PHM will better reproduce cycle loadings as well as the strength at strains above the maximum stress.

## Acknowledgments

The author wishes to thank D. Porat for carefully performing all experiments described in this paper, and to Y. Partom, and A. Malka-Markovitch for reading the manuscript and providing valuable advice.